\documentclass[12pt]{emulateapj}
\slugcomment{{\sc Accepted for publication in The Astrophysical Journal:} November 8, 2010}

\def\etl{et al.}
\def\eg{e.g., }

\newcommand{\sdeg}{\ensuremath{{\rm deg}^2}}
\newcommand{\mi}{\ensuremath{M_{^{0.3}i}-5\log_{10}h}}
\newcommand{\mr}{\ensuremath{M_{^{0.1}r}-5\log_{10}h}}
\newcommand{\mgmi}{\ensuremath{M_{^{0.3}g}-M_{^{0.3}i}}}
\newcommand{\monemtwo}{\ensuremath{M_{^{0.3}[8.0]}-M_{^{0.3}[3.6]}}}
\newcommand{\chone}{\ensuremath{3.6~\mu{\rm m}}}
\newcommand{\chtwo}{\ensuremath{4.5~\mu{\rm m}}}
\newcommand{\chthree}{\ensuremath{5.8~\mu{\rm m}}}
\newcommand{\chfour}{\ensuremath{8.0~\mu{\rm m}}}
\newcommand{\mipschone}{\ensuremath{24~\mu{\rm m}}}

\usepackage{natbib}

\begin{document}
\shorttitle{PRIMUS: Obscured Star Formation}
\shortauthors{Zhu \etl}
\title {PRIMUS: Obscured Star Formation on the Red Sequence}

\author{
 Guangtun Zhu\altaffilmark{1},
 Michael R. Blanton\altaffilmark{1},
 Scott M. Burles\altaffilmark{2},
 Alison L. Coil\altaffilmark{3,7},
 Richard J. Cool\altaffilmark{4,8},
 Daniel J. Eisenstein\altaffilmark{5},
 John Moustakas\altaffilmark{3},
 Kenneth C. Wong\altaffilmark{6},
 and James Aird\altaffilmark{3}
} 
\altaffiltext{1}{Center for Cosmology and Particle Physics, Department of 
Physics, New York University, 4 Washington Place, New York, NY 10003; 
gz323@nyu.edu}
\altaffiltext{2}{D.E. Shaw \& Co., L.P., 20400 Stevens Creek Blvd., Suite 
850, Cupertino, CA 95014}
\altaffiltext{3}{Department of Physics, Center for Astrophysics and Space 
Sciences, University of California, 9500 Gilman Dr., La Jolla, San Diego, 
CA 92093}
\altaffiltext{4}{Department of Astrophysical Sciences, Princeton University, 
Peyton Hall, Princeton, NJ 08544}
\altaffiltext{5}{Department of Astronomy, Harvard University, 60 Garden Street,
Cambridge, MA 02138}
\altaffiltext{6}{Steward Observatory, University of Arizona, 933 North Cherry 
Avenue, Tuscon, AZ 85721}
\altaffiltext{7}{Alfred P. Sloan Foundation Fellow}
\altaffiltext{8}{Hubble Fellow and Carnegie-Princeton Fellow}

\begin{abstract}
 We quantify the fraction of galaxies at moderate redshifts
 ($0.1<z<0.5$) that appear red-and-dead in the optical, but in fact
 contain obscured star formation detectable in the infrared (IR), with the
 PRIsm MUlti-object Survey (PRIMUS). PRIMUS has measured $\sim120,000$ robust
 redshifts with a precision of $\sigma_z/(1+z)\sim0.5\%$ over $9.1~\sdeg$ of 
 the sky to the depth of $i\sim23$ (AB), up to redshift $z\sim1$.
 We specifically targeted $6.7~\sdeg$ fields with existing deep
 IR imaging from the {\it Spitzer Space Telescope} from the
 SWIRE and S-COSMOS surveys.  We select in these fields an $i$ band
 flux-limited sample ($i<20$ mag in the SWIRE fields and $i<21$ mag in
 the S-COSMOS field) of $3310$ red-sequence galaxies at $0.1<z<0.5$
 for which we can reliably classify obscured star-forming and
 quiescent galaxies using IR color.  Our sample constitutes the
 largest galaxy sample at intermediate redshift to study obscured star
 formation on the red sequence, and we present the first quantitative
 analysis of the fraction of obscured star-forming galaxies as a
 function of luminosity.
 We find that on average, at $L \sim L^{\ast}$, about $15\%$ of
 red-sequence galaxies have IR colors consistent with star-forming
 galaxies.  The percentage of obscured star-forming galaxies increases
 by $\sim8\%$ per mag with decreasing luminosity from the highest
 luminosities to $L\sim 0.2L^\ast$.
 Our results suggest that a significant fraction of red-sequence
 galaxies have ongoing star formation and that galaxy evolution studies
 based on optical color therefore need to account for this
 complication.
\end{abstract}

\keywords{galaxies: fundamental parameters (classification, colors, 
  luminosities, masses, radii, etc.) ---
  galaxies: distances and redshifts --
  galaxies: evolution --
  stars: formation --
  dust, extinction}

\section {Introduction}

 Our understanding of the formation and evolution of galaxies has improved 
 tremendously the last few decades as a result of multiple redshift surveys 
 \citep[\eg][]{davis82a, york00a, colless01a, wolf03a, davis03a, lilly07a, 
 garilli08a}.  These surveys have demonstrated that the distribution of 
 galaxy populations is bimodal in the optical color-magnitude space up 
 to $z\sim2.5$ \citep[\eg][among others]{strateva01a, blanton03b, 
 baldry04a, balogh04a, bell04a, willmer06a, wuyts07a, williams09a, brammer09a}. 
 The underlying physical reason is the short lifetime of massive stars.
 Once star formation stops in a galaxy, hot massive stars quickly
 die, the integrated spectral energy distribution (SED) of the
 galaxy becomes 
 dominated by cool giants, and its optical color turns from blue 
 to red.  The bimodality of blue cloud and red sequence in the 
 color-magnitude diagram therefore approximately translates into that of 
 star-forming (SF) and quiescent galaxies.

 Recent studies of the mass and luminosity functions of red-sequence
 galaxies at different redshifts show that the stellar mass density on
 the red sequence has increased by a factor of $\sim 2$ overall since
 $z\sim1$ \citep[\eg][]{bell04a, faber07a}. This build-up could result
 from some combination of migration from the blue cloud (due to
 mergers, feedback or other mechanisms for turning off star formation)
 and gas-poor, ``dry'' merging from smaller-mass red-sequence galaxies
 \citep[\eg][]{bell04a, bell06b, bell06a}.  Other studies
 \citep[\eg][]{cimatti06a,brown07a,cool08a} find that the growth of
 the red-sequence populations depends strongly on luminosity: whereas
 many $L^{\ast}$ red-sequence galaxies assembled in the time since
 $z\sim1$, at the massive end ($L \gtrsim 3L^{\ast}$) the mass and
 luminosity functions are consistent with passive evolution. They
 therefore argue that merging (dry or wet) does not dominate the
 evolution of massive red galaxies since $z\sim1$. 

 One complication that current studies of red-sequence galaxies ignore
 is that optically red galaxies can have {\it in situ}, but dust-obscured
 star formation.
 Because of this dust-obscuration, optical color can be a misleading
 proxy for star formation.  Dust extinction reshapes the SED of a
 galaxy, absorbing photons at high energy (\eg ultraviolet, UV) and
 re-emitting the energy at longer wavelengths (\eg infrared, IR).
 Therefore the red sequence consists of both truly quiescent but
 also obscured SF galaxies; it is extremely important to quantify the
 composition of the red sequence to obtain an accurate understanding
 of galaxy evolution.

 There are several ways to investigate the contribution of obscured SF galaxies 
 on the red sequence.  Assuming inclination is well-correlated with dust 
 extinction along the line-of-sight, \citet[][see also \citealt{shao07a, 
 unterborn08a, bailin08b}]{maller09a} find the ratio of red-to-blue galaxies 
 changes from 1:1 to 1:2 when moving from observed to inclination-corrected 
 color, for local galaxies in the Sloan Digital Sky Survey (SDSS,
 \citealt{york00a}).
Using deep imaging from the {\it Hubble Space Telescope} 
 ({\em HST}) from the Galaxy Evolution from Morphology and SED project 
 \citep[GEMS, ][]{rix04a} and photometric redshifts from the COMBO-17 
 survey \citep{wolf03a}, \citet{bell04b} find that less than $13\%$ of 
 red-sequence galaxies are edge-on dusty disk galaxies at $z\sim0.7$.
 One can also correct dust extinction by fitting multi-wavelength data
 or SEDs with stellar population synthesis models and allowing dust 
 extinction to be a free parameter.  Using this method on COMBO-17
 data, \citet{wolf05a} 
 find a rich component of dusty SF galaxies contaminating the red sequence 
 in the cluster Abell $901/902$ at $z\sim0.17$.  The mid-IR (MIR) emission 
 from dust offers another method for selecting obscured SF 
 galaxies, though the emission can also originate from strong active galactic 
 nuclei (AGNs).  \citet[][see also \citealt{coia05a, davoodi06a, rodighiero07a, 
 saintonge08a}]{brand09a} discover $\sim 10\%$ of red-sequence galaxies 
 at $0.15 \leq z \leq 0.3$ exhibit strong MIR \mipschone~emission
 with $f_{24}>300~\mu$Jy.
 We can also use IR color to separate SF and quiescent galaxies.
 At wavelengths $<10~\mu$m, warm dust emission and strong polycyclic 
 aromatic hydrocarbon (PAH) features dominate the SED of SF galaxies,
 producing much redder  IR colors than expected from quiescent galaxies.
 \citet{brand09a} use this method and find that, among the $\sim10\%$ of 
 red-sequence galaxies with $f_{24}>300~\mu$Jy in their sample, 
 about $64\%$ have IR colors consistent with SF galaxies, with the rest 
 consistent with quiescent galaxies ($31\%$) and AGNs ($5\%$).

 Quantitative studies of obscured star formation at intermediate 
 and high redshifts are usually limited to small samples. For example, 
 \citet{wolf05a} select $462$ red-sequence galaxies in a specific environment,
 the Abell $901$/$902$ super-cluster system.
 Even though deep multi-wavelength data including
 space-based IR imaging are now available across a large area of the sky, 
 the lack of corresponding large redshift surveys hampers the systematic 
 quantitative studies of the contribution of obscured SF galaxies on the red 
 sequence. We have recently completed such a redshift survey, the PRIsm 
 MUlti-object Survey 
 (PRIMUS\footnote{http://cass.ucsd.edu/$\sim$acoil/primus/}),
 to provide the redshifts required to fully realize the science potential of
 the existing deep multi-wavelength imaging data.

 PRIMUS has measured $\sim 120,000$ robust redshifts 
 over $9.1~\sdeg$ of the sky 
 to the depth of $i\sim23$ (AB) up to redshift $z\sim 1$.
 We specifically targeted areas where there exist deep multi-wavelength 
 imaging data, including IR data from the {\it Spitzer Space Telescope} 
 \citep[][]{werner04a}, making our dataset ideal for the study of the obscured 
 star formation on the red sequence.  Using the IR data, in conjunction with 
 the redshifts from PRIMUS, we here present the first quantitative study 
 of the fraction of obscured SF galaxies on the red sequence
 as a function of luminosity at redshifts $0.1<z<0.5$. 

 This paper proceeds as follows.
 In Section \ref{sec:data}, we briefly introduce PRIMUS and describe 
 the {\it Spitzer} data we use. In Section \ref{sec:selection}, we describe 
 in detail the classification scheme for AGNs, star-forming galaxies, and 
 quiescent galaxies.  We present the results in Section \ref{sec:results}.  
 In Section \ref{sec:discussion}, we discuss the composition of the red 
 sequence.
 We summarize our principal conclusions in Section \ref{sec:summary}.

 Throughout this work, we adopt a $\Lambda$CDM cosmological model with 
 $\Omega_{\mathrm{m}}=0.3$, $\Omega_\Lambda=0.7$ and
 $H_0=100h~\mathrm{km~s^{-1}~Mpc^{-1}}$.  
 If not specifically mentioned, all magnitudes are on the AB system.

\section{Data}\label{sec:data}

\subsection{PRIMUS}\label{sec:dataprimus}

 PRIMUS is a unique low-resolution spectroscopic intermediate redshift survey.
 We conducted the survey using the IMACS instrument on the
 Magellan I (Baade) Telescope at Las Campanas Observatory.
 With a prism and slitmasks, we obtained low-resolution spectra for
 $\sim 300,000$ objects to the depth of $i\sim23$ out to redshift $z\sim1$ 
 in $39$ nights, covering $9.1~{\rm deg}^2$ of the sky.
 The low-resolution of the prism ($R \sim 30-150$) allows us 
 to observe $\sim 2500$ objects in one single mask with a field of view 
 $0.18~\sdeg$.  By carefully fitting the low-resolution spectra with galaxy 
 spectral templates, we have obtained $\sim 120,000$ robust redshifts, 
 to the precision of $\sigma_z/(1+z) \sim 0.5\%$. PRIMUS therefore is the 
 largest faint galaxy redshift survey performed to date.

 When designing the survey, we specifically targeted fields with existing deep 
 multi-wavelength imaging, including optical imaging from various 
 ground-based deep surveys, near-UV (NUV) and far-UV (FUV) 
 from the Galaxy Evolution Explorer ({\em GALEX}, \citealt{martin05a}),
 IR from the {\it Spitzer Space Telescope},
 and $X$-ray from {\it Chandra} and/or {\em XMM} telescopes. 

 We created homogeneous band-merged catalogs across all our fields. 
 With the redshifts and the band-merged catalogs, 
 we derived absolute magnitudes using the {\tt kcorrect} package 
 \citep[v4.1.4][]{blanton07a}.
 Because we focus on redshifts $0.1<z<0.5$, here we use the absolute magnitudes 
 at $ugriz$ bands shifted blueward to $z=0.3$ \citep[\eg][]{blanton03b}. 
 We denote these bands with a preceding superscript $0.3$, e.g., $^{0.3}i$.
 For a galaxy exactly at redshift $0.3$ with apparent magnitude $i$, the 
 $K$-correction from $i$ to $M_{^{0.3}i}$ is simply $-2.5\log_{10}(1+z)$
 independent of galaxy SED.  This filter choice therefore minimizes the error 
 in the $K$-corrections.

 When fitting for the redshift, we also fit the spectra with stellar and 
 broad-line AGN spectral templates and classify objects into stars, 
 AGNs, and galaxies based upon the $\chi^2$ of the best fits. 
 In $\sim 5.5~\sdeg$ of the $9.1~\sdeg$ PRIMUS 
 fields, we have $X$-ray data from {\it Chandra} and/or {\em XMM} telescopes.
 Here we select our sample only from objects that are classified as galaxies and
 do not have matched $X$-ray detections. 

 For more details on the PRIMUS survey, we refer the reader to the survey papers
 \citep[][Cool \etl, in prep.]{coil10a}.

\subsection{\it Spitzer Space Telescope}

 In $\sim 7.2~\sdeg$ of the $9.1~\sdeg$ fields covered by PRIMUS
 we have IR imaging data from The Infrared Array Camera 
 (IRAC, \citealt{fazio04a}) and The Multi-band Imaging Photometer 
 (MIPS, \citealt{rieke04a}) onboard {\it Spitzer}.  IRAC is a 
 four-channel instrument that provides simultaneous broadband images 
 at $3.6$, $4.5$, $5.8$ and \chfour~with unprecedented sensitivity.  
 MIPS provides images at longer wavelengths, $24$, $70$ and $160~\mu$m 
 with $5\sigma$ rms depths of $0.3$ mJy, $25$ mJy and $150$ mJy, respectively.
 Here we use the $\sim6.7~\sdeg$ that has coverage from all four channels
 of IRAC and $24~\mu$m band of MIPS. For $\sim0.5~\sdeg$ of the DEEP2 
 \citep{davis03a} field with the IRAC imaging, 
 only the  \chone{} and \chtwo{} imaging is available; we omit
 this field, because we require all four channels for this paper.
 The data we use consist of two Legacy Programs of {\it Spitzer},
  $5.7~\sdeg$ in three fields of {\it Spitzer} Wide-area Infrared 
  Extragalactic Survey
 (SWIRE\footnote{http://swire.ipac.caltech.edu/swire/swire.html}; 
 \citealt{lonsdale03a}), and  $1~\sdeg$ in the {\it Spitzer} Deep Survey of the {\em HST} 
 COSMOS 2-Degree ACS Field (S-COSMOS\footnote{http://irsa.ipac.caltech.edu/data/SPITZER/S-COSMOS/; 
 COSMOS represents the Cosmological Evolution Survey and ACS is the Advanced 
 Camera for Surveys onboard {\em HST}.}; \citealt{sanders07a}).
 We use the data release $2$ and $3$ of the SWIRE survey, and the public catalog 
 in the NASA/IPAC Infrared Science Archive for the S-COSMOS survey.
 In most of the SWIRE fields, to at least $5\sigma$ depths, the flux limits
 in the four IRAC channels are $10$, $10$, $43$, and $40~\mu$Jy 
 ($21.4$, $21.4$, $19.8$ and $19.9$ AB mag\footnote{We use AB magnitudes 
 throughout this work but also give Vega magnitudes when necessary. 
 The conversion factors (AB$-$Vega) for IRAC $3.6$, $4.5$, $5.8$, 
 and \chfour~are $2.79$, $3.26$, $3.73$, and $4.40$ mag, respectively.}). 
 The flux limits in the S-COSMOS field are $0.9$, $1.7$, $11.3$, and 
 $14.6~\mu$Jy ($24.0$, $23.3$, $21.3$ and $21.0$ AB mag), in the four 
 IRAC channels, respectively.

 At wavelengths $ < 10~\mu$m, warm dust emission and strong PAH features, e.g, 
 $3.3$, $6.2$, $7.7$ and $8.6~\mu$m, dominate the SED
 of a typical SF galaxy \citep{li01a, smith07a}.  Moving to 
 higher redshift, the $6.2$, $7.7$ and $8.6~\mu$m PAH emission bands shift 
 through the IRAC \chfour~channel, producing much redder IRAC colors than 
 expected from a typical quiescent early-type galaxy. Red-sequence 
 galaxies whose star formation is obscured by dust
 have IRAC colors consistent with normal SF galaxies.  
 Strong AGNs, on the other hand, occupy a different locus from 
 regular SF and quiescent galaxies in the IRAC color-color diagram 
 \citep[][]{eisenhardt04a, lacy04a, stern05a}.
 The IRAC colors therefore are extremely convenient and powerful for 
 classification of AGNs, SF galaxies, and quiescent galaxies.
 We discuss this in detail in Section \ref{sec:selection}.

\section{Source Selection}\label{sec:selection}

\subsection{Flux-limited sample in $i$ band}\label{sec:ilimit}

\begin{figure}
\epsscale{1.2}
\plotone{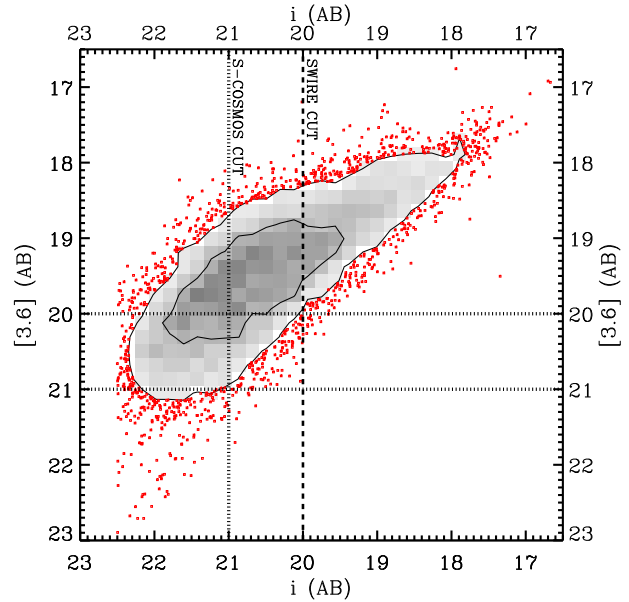}
\caption{Selection of the $i$ band flux-limited sample.
 We present $[3.6]$ vs. $i$ for red-sequence galaxies.
 The detection limits in the \chfour~channel are 
 $19.9$ ($21.0$) mag in the SWIRE (S-COSMOS) fields, effectively much 
 brighter than the limits in the \chone~channel. 
 Many \chone~detections are not detected in the \chfour~channel. 
 We can only use the IR color $[3.6]-[8.0]$ to 
 perform reliable star-forming and quiescent galaxy classification for 
 objects with bright \chone~fluxes with $[3.6]\lesssim20$ ($21$) mag.
 We show here that most of the red-sequence galaxies with $i<20$ ($21$) mag
 have $[3.6]<20$ ($21$) mag.
 We therefore only select galaxies with $i$ brighter than $20$ ($21$) mag 
 in the SWIRE (S-COSMOS) fields.  In Section \ref{sec:sfclass}, we show that 
 these cuts enable us to classify almost all non-detections in \chfour~channel 
 as quiescent galaxies in this flux-limited red-sequence sample.
}
\label{fig:chonei}
\end{figure}

\begin{figure}
\epsscale{1.2}
\plotone{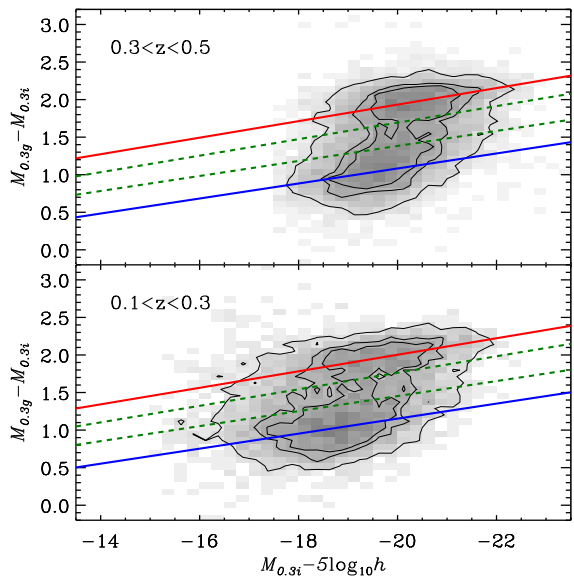}
\caption{Color-magnitude diagram of galaxies with $i<22.5$. 
 We use the absolute magnitudes in $g$ and $i$ bands shifted blueward to $z=0.3$ 
 ($^{0.3}g$ and $^{0.3}i$) to minimize the errors in $K$-corrections.
 We show the diagram in two redshift bins: $0.1<z<0.3$ ({\it lower panel}) 
 and $0.3<z<0.5$ ({\it upper panel}).
 The red solid lines and the blue solid lines indicate the positions 
 of the red sequence and the blue cloud. 
 The two green dashed lines represent the selection 
 criteria (Equation (1) and (2)) we adopt to select red-sequence and blue-cloud 
 galaxies, at median redshift in each bin. 
 Galaxies between them are identified as green-valley galaxies.
 At different redshifts, we fix the slope of the criteria but allow the intercept
 to evolve (redden) by $0.4$ per unit redshift.
}
\label{fig:cmd}
\end{figure}

 In order to use IR color to identify SF galaxies,
 we restrict our sample to the PRIMUS fields with IR coverage ($6.7~\sdeg$).
 We also only select our sample from galaxies with the highest confidence 
 ({\tt Q} $=4$, see Cool \etl~in prep. for details on the confidence level 
 {\tt Q}).

 We also need to select appropriate flux limits in the $i$ band and the 
 \chone~band, which are necessarily related to the depth of the coverage 
 in the \chfour~band.  The flux limits used in the \chfour~channel are 
 $19.9$ mag in the SWIRE fields and $21.0$ mag in the S-COSMOS field, 
 effectively much brighter than those in the \chone~channel 
 ($21.4$ and $24.0$ mag). Thus, in order to place interesting limits on the 
 specific star-formation rate (SSFR)
 using the $[3.6]-[8.0]$ color, we must restrict our sample to bright 
 enough galaxies.
 In Section \ref{sec:sfclass}, we will show that galaxies with
 SSFR of $\sim 10^{-10}~{\rm yr}^{-1}$ have colors $[3.6]-[8.0]\gtrsim0$ 
 and quiescent galaxies have 
 $[3.6]-[8.0]\lesssim0$.  Our definition of ``star-forming'' will be 
 loosely related to this SSFR, and this color cut defines how bright our 
 sample needs to be.  We thus require the upper limits of the $[3.6]-[8.0]$ 
 colors of the \chfour{} non-detections to be $\lesssim 0$ so that we can 
 reliably classify them into quiescent galaxies, or in other words that
 the $[3.6]$ magnitudes should be brighter than $\sim20$ ($21$) mag 
 in the SWIRE (S-COSMOS) fields. 
 
 We achieve this by applying appropriate limits in the $i$ band.  
 In Figure \ref{fig:chonei}, we present the $[3.6]$ versus $i$ for the 
 red-sequence galaxies (defined in the next section), where we have 
 shown $[3.6]=20$ and $21$ mag with two horizontal lines. We see that for 
 most of the red-sequence galaxies with $i<20$ ($21$) mag, they have 
 \chone~magnitudes $[3.6]<20$ ($21$). We therefore define a flux-limited 
 sample with $i<20$ and $i<21$, in the SWIRE and S-COSMOS fields, 
 respectively.  We will further justify these cuts in Section \ref{sec:sfclass},
 where we explicitly demonstrate that after these cuts, virtually all 
 \chfour~non-detections are quiescent galaxies.

 Beyond redshift $0.5$, the $6.2~\mu$m PAH feature shifts out of the 
 \chfour~channel and we can not use the $[3.6]-[8.0]$ color to reliably 
 separate SF and quiescent galaxies.  We also do not have many galaxies 
 below redshift $0.1$ in the PRIMUS fields due to the small volume.  
 Therefore, we focus our analysis on the redshift range $0.1<z<0.5$.  
 We will come back to this in Section \ref{sec:sfclass}.  In total, 
 the flux-limited sample within $0.1<z<0.5$ includes $7258$ galaxies.

\subsection{Red sequence, blue cloud, and green valley}\label{sec:cmd}

\begin{figure}
\epsscale{1.2}
\plotone{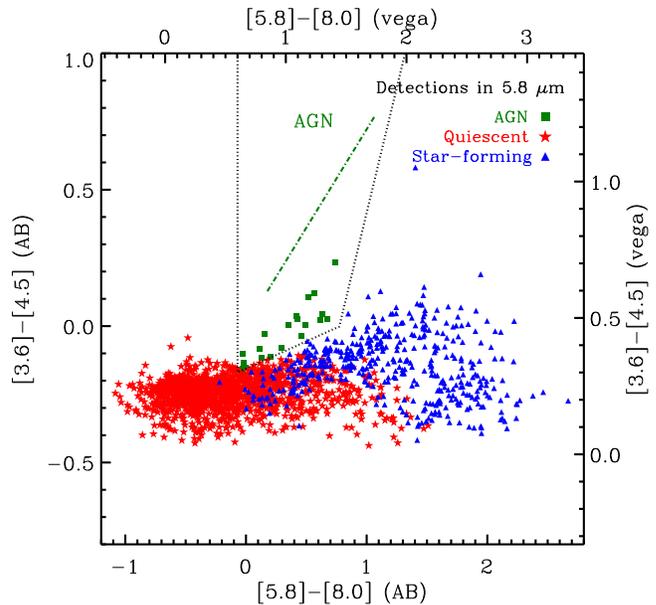}
\caption{AGN classification for detections in both 
 the \chthree~and the \chfour~channels in the flux-limited red-sequence sample.
 We show the IRAC color-color diagram with $[3.6]-[4.5]$ vs. $[5.8]-[8.0]$. 
 The dotted wedge shows the criteria for AGN selection suggested by 
 \citet{stern05a}. The dot-dashed line represents the colors of 
 power-law spectra $f(\nu)=\nu^{-\alpha}$ of AGNs with the index 
 $\alpha$ ranging from $0.5$ to $3.0$.  Galaxies that fall into the wedge 
 are identified as AGNs. For non-AGN galaxies, we use the 
 $[3.6]-[8.0]$ color to classify them into star-forming galaxies and quiescent 
 galaxies (Section \ref{sec:sfclass}).  
 We show the SF galaxies in blue filled triangles and
 the quiescent galaxies in red filled stars.
}
\label{fig:iraccolor}
\end{figure}

 To define the red sequence, blue cloud, and green valley, we use all galaxies 
 with $i<22.5$ with secure redshifts ({\tt Q} $=4$) in PRIMUS.  
 PRIMUS targeted in $i$ band unless there is no $i$ band imaging available,
 in which case we used $R$ band.
 We thus choose to use the magnitude \mi, and the color \mgmi~when
 defining samples based on optical color.
 At each luminosity, we fit the color distribution with a two-gaussian model. 
 We then linearly fit the two sets of peaks to obtain the 
 slopes and intercepts of the red sequence and the blue cloud.  
 To define the edges of green valley, we move the fitted line for the 
 red sequence (blue cloud) blueward (redward) until it includes $60\%$ 
 more galaxies above (below) the line.  This method allows us to conservatively 
 select red-sequence and blue-cloud galaxies, without contamination from 
 green-valley galaxies.  Due to passive evolution, the colors of galaxies 
 at lower redshift are redder than those at higher redshift.  
 Therefore, we keep the slope 
 fixed throughout the redshift range, but allow the intercept to evolve 
 linearly in redshift.  By fitting the color-magnitude diagram 
 at different redshifts, we measure the color reddening is $\sim 0.4$ 
 per unit redshift in \mgmi.

 Finally, the criteria we adopt are as follows.
 We require a red-sequence galaxy to have the color:
 \begin{eqnarray}
  \mgmi > & & \nonumber \\
 & 1.73 - 0.11 (\mi + 20) & \nonumber \\
 & - 0.4 (z-0.3) \mathrm{,} & 
 \end{eqnarray}
 \noindent and a blue-cloud galaxy to have the color:
 \begin{eqnarray}
 \mgmi <  & & \nonumber\\
 & 1.40 - 0.10 (\mi + 20) & \nonumber \\
 & - 0.4 (z-0.3) \mathrm{.} & 
 \end{eqnarray}
 \noindent Between these two criteria, we identify objects as green-valley 
 galaxies.  As an example, in Figure \ref{fig:cmd}, we show the 
 color-magnitude diagram of all galaxies with $i<22.5$ at $0.3<z<0.5$ 
 ({\it upper panel}) and at $0.1<z<0.3$ ({\it lower panel}). 
 The two solid lines in each panel represent the positions of the 
 red sequence and the blue cloud, and the dashed lines indicate the 
 edges of the green valley, i.e., the criteria defined by Equation 
 (1) and (2) at median redshift in each bin. 

 Finally, in the flux-limited sample, we have $3310$ red-sequence galaxies, 
 $2497$ blue-cloud galaxies and $1451$ green-valley galaxies.

\subsection{AGN identification}\label{sec:agnid}

\begin{figure}
\epsscale{1.2}
\plotone{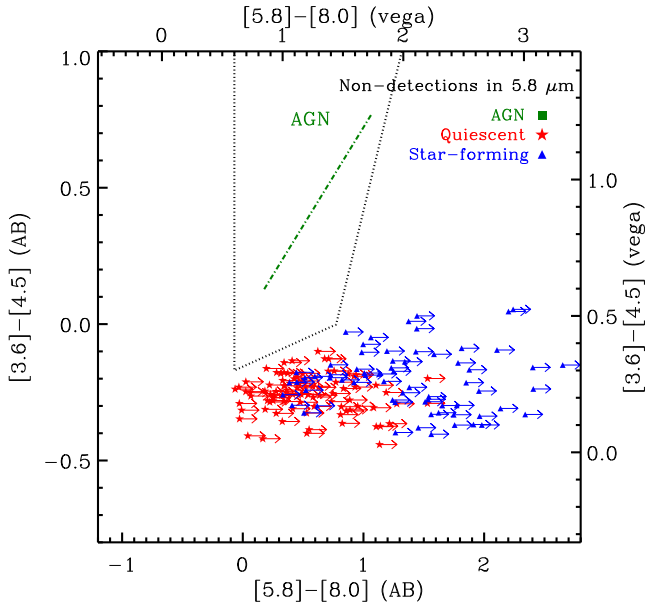}
\caption{AGN classification for non-detections in the \chthree~channel that are 
 detected in the \chfour~channel in the flux-limited red-sequence sample. 
 We use the flux limits of $[5.8]<19.8$ ($21.0$) mag in the SWIRE 
 (S-COSMOS) fields to calculate the lower limits of their $[5.8]-[8.0]$ colors. 
 We can safely treat all the non-detections as non-AGNs.
}
\label{fig:iraccolorchthree}
\end{figure}

 For the purposes of this paper, we are primarily interested in
 star-forming galaxies, and attempt to identify and exclude the few
 AGNs in the sample. To do so, we use a combination of the PRIMUS
 spectra, matching $X$-ray data, and IR colors of the galaxies.

 As stated in Section \ref{sec:dataprimus}, we only select objects 
 classified as galaxies and did not include those classified as 
 broad-line AGNs in the PRIMUS spectroscopy.  For $X$-ray AGNs, we 
 exclude galaxies with matched $X$-ray detections in the $4.3~\sdeg$ 
 of $X$-ray coverage from {\it Chandra} and/or {\em XMM} telescopes
 (out of our full $6.7~\sdeg$).  However, neither technique can exclude 
 heavily-obscured AGNs effectively. 

 Finally, IR colors can be used to exclude strong AGNs.
 In the IRAC color-color diagram with $[3.6]-[4.5]$ versus $[5.8]-[8.0]$, 
 AGNs occupy a different locus from normal galaxies, either SF or 
 quiescent \citep[][]{lacy04a, stern05a}. This is because the SED of an 
 AGN in the IR can be well-represented by a power law, 
 while that of a quiescent 
 galaxy approximately follows Rayleigh-Jeans law and that of a SF galaxy
 is dominated by warm dust and PAH emission.  
 To identify AGNs, we adopt the empirical criteria 
 suggested by \citet{stern05a}. These criteria may 
 omit AGNs at high redshift, but are adequate 
 for our purpose since our sample is at $z<0.5$. 
 In Figure \ref{fig:iraccolor}, we show the IRAC color-color diagram for 
 all the detections in both the $5.8$ and $8.0$ channels in the flux-limited 
 red-sequence sample,
 with \citet{stern05a} criteria as the dotted wedge.
 Objects within the wedge are classified as AGNs (green filled squares). 
 We also show SF and quiescent galaxies among the non-AGNs, defined in the 
 next section, with blue filled triangles and red filled stars, respectively.
 The green dot-dashed line represents the colors of AGNs with power-law spectra 
  $f(\nu)=\nu^{-\alpha}$, with $\alpha$ ranging from $0.5$ to $3.0$.

 The \citet{stern05a} method to identify AGNs requires detections in all four 
 channels.  In the flux-limited red-sequence sample, a large fraction 
 ($1288$ of $3310$) are not detected in the \chfour~channel.  Nevertheless, 
 we will show in the next section that we can safely classify them as 
 quiescent galaxies. Among those ($2022$ of $3310$) detected in the 
 \chfour~channel, $154$ do not have matched detections in the \chthree~channel 
 due to the relatively bright flux limits. For these objects, we use the 
 flux limits of $[5.8]<19.8$ ($21.3$) mag in the SWIRE (S-COSMOS) fields 
 to calculate the lower limits of their $[5.8]-[8.0]$ colors.  We show the 
 results in Figure \ref{fig:iraccolorchthree}.  The colors of all 
 \chthree~non-detections are unambiguously outside the AGN wedge and we 
 therefore can safely classify them as non-AGNs and perform further SF and 
 quiescent galaxy classification in the next section.

\subsection{Classification of star-forming and quiescent 
galaxies}\label{sec:sfclass}

\begin{figure}
\epsscale{1.2}
\plotone{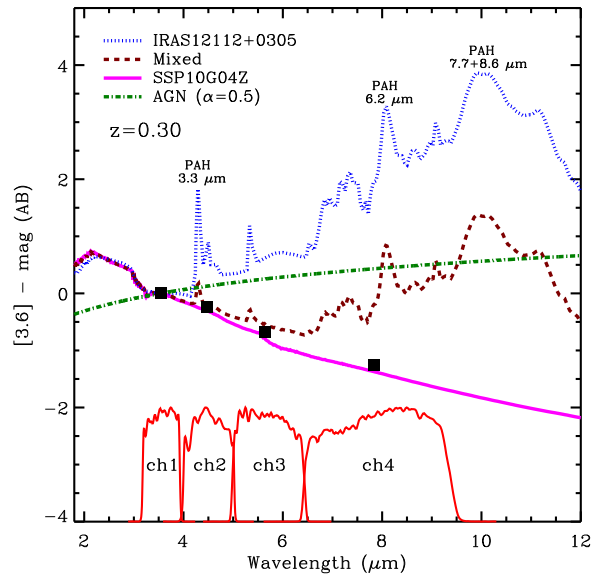}
\caption{Spectral templates shifted to $z=0.3$, 
 in AB magnitudes, normalized to 
 have the same magnitude zeropoint in the \chone~channel. 
 We use the template of IRAS$12112+0305$ from \citet{rieke09a} to represent 
 strong star-forming galaxies.  For early-type galaxies without star formation 
 and dust, we use a simple stellar population (SSP) with an age of $10$ Gyr 
 and metallicity of $0.4Z_\odot$ (SSP10G04Z) from the \citet{bruzual03a} 
 models. This template is approximately the average of a variety of SSPs 
 with age ranging from $5$ to $10$ Gyr and metallicity of $0.4Z_\odot$ 
 and $1.0Z_\odot$.  We show the average magnitudes in IRAC 
 channels of these SSPs in black filled squares.  To classify star-forming 
 and quiescent galaxies, we define a demarcation template ({\tt Mixed}) 
 by linearly combining the two templates. Normalized at $2~\mu$m, 
 the {\tt Mixed} template is the sum of $10\%$ of IRAS$12112+0305$ and 
 $90\%$ of SSP10G04Z, and represents a galaxy with SSFR 
 $\sim 10^{-10}~{\rm yr}^{-1}$.  The green dot-dashed line represents a 
 power-law spectrum $f(\nu) = \nu^{-\alpha}$ of an AGN with $\alpha=0.5$. 
 We also show the response curves of IRAC channels on the bottom.}
\label{fig:template}
\end{figure}

\begin{figure}
\epsscale{1.2}
\plotone{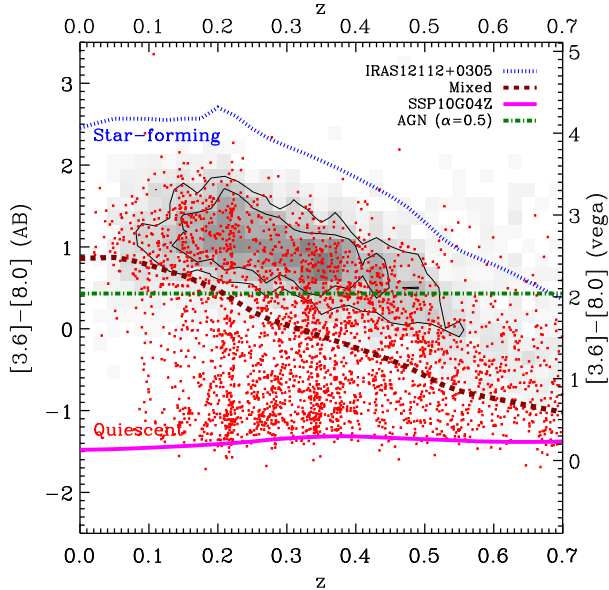}
\caption{IR Color $[3.6]-[8.0]$ of the spectral templates in 
 Figure \ref{fig:template}, as a function of redshift.
 The various lines represent the colors of all the spectral templates 
 in the same colors and line styles as in Figure \ref{fig:template}.  For 
 comparison, we also show all the \chfour~detections in the blue-cloud 
 (gray scale) and red-sequence galaxies (red dots) with $i<22.5$ at $z<0.7$.  
 We classify galaxies above the {\tt Mixed} template (the brown dashed line) 
 as star-forming galaxies, if they are not in the AGN wedge 
 in the IRAC color-color diagram (Figure \ref{fig:iraccolor}),  
 At $z>0.5$, the $6.2~\mu$m PAH feature shifts out of the \chfour~channel,
 causing the colors of star-forming and quiescent galaxies to blend into each 
 other. 
}
\label{fig:iracz}
\end{figure}

 To separate SF and quiescent galaxies, we combine representative SED 
 templates of strong SF galaxies and typical quiescent early-type galaxies.
 For strong SF galaxies, we choose the template of IRAS$12112+0305$ from 
 \citet{rieke09a}, an ultra-luminous IR galaxy (ULIRG) with significant
 ongoing star formation \citep[\eg][]{armus07a}.  For typical quiescent 
 early-type galaxies, we use a simple stellar population (SSP) with an 
 age of $10$ Gyr and metallicity of $0.4Z_\odot$ from the \citet{bruzual03a} 
 models, with the Padova 1994 library of stellar evolution tracks and the 
 \citet{chabrier03a} Initial Mass Function (IMF).
 We denote this template as SSP10G04Z.  In Figure \ref{fig:template}, 
 we show these templates, shifted to $z=0.3$, within
 the wavelength range $2$ to $12~\mu$m. We also show a power-law spectrum
 $f(\nu)=\nu^{-\alpha}$ with $\alpha=0.5$ to represent a typical AGN. 
 Note we have normalized all the SEDs to have the same magnitude zeropoint 
 in the \chone~channel for clarity.  The IRAS$12112+0305$ template is 
 dominated by warm dust continuum and strong PAH features.  Meanwhile, 
 the SSP template roughly follows the Rayleigh-Jeans law 
 $f(\lambda) \propto \lambda^{-4}$.  This template is similar to the 
 average of a variety of SSPs with age ranging from $5$ Gyr to $15$ Gyr 
 and metallicity of $0.4Z_\odot$ and $1.0Z_\odot$.  We show the average 
 magnitudes of these SSPs in the IRAC channels in black filled squares.

 To define a demarcation template separating SF and quiescent galaxies,
 we linearly combine the templates of IRAS$12112+0305$ and SSP10G04Z.
 We first normalize them to have the same flux at $2~\mu$m (roughly $K$-band),
 and then define the demarcation mixed template ({\tt Mixed} hereafter) to 
 be the sum of $10\%$ of IRAS$12112+0305$ and $90\%$ of SSP10G04Z. 
 This {\tt Mixed} template represents a galaxy with SSFR 
 $\sim 10^{-10}~{\rm yr}^{-1}$.  We show the template with the brown dashed 
 line in Figure \ref{fig:template}.

\begin{figure}
\epsscale{1.2}
\plotone{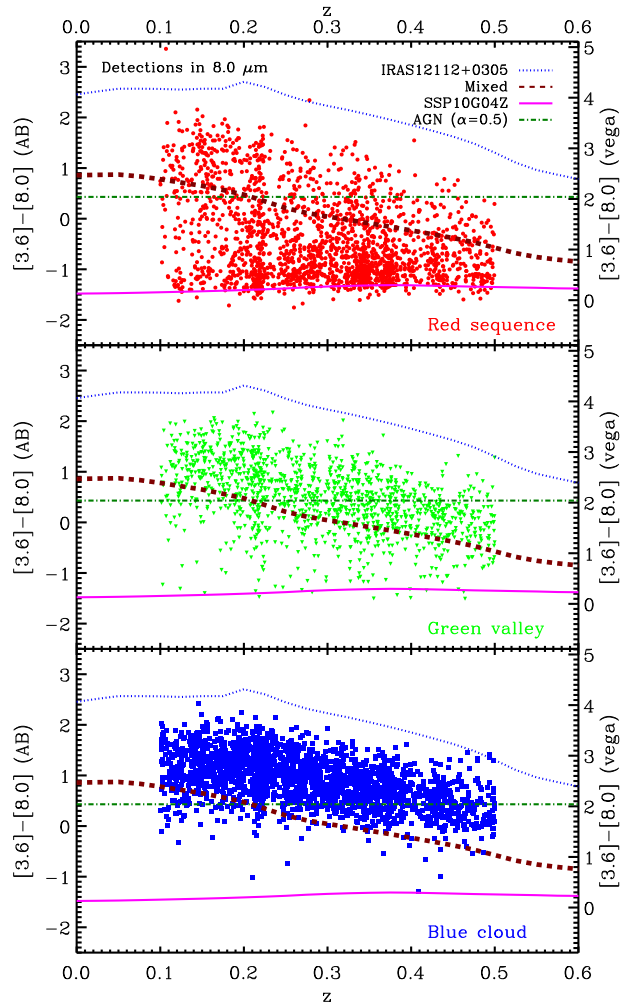}
\caption{IR Color $[3.6]-[8.0]$ vs. z for all \chfour~detections in the 
 flux-limited sample ($i<20$ mag in the SWIRE fields and $i<21$ mag in 
 the S-COSMOS field).  {\it Top panel}: red-sequence galaxies. Many are 
 scattered into the star-forming region (above the {\tt Mixed} template).
 {\it Middle panel}: green-valley galaxies. Most of them appear in the 
 star-forming region.
 {\it Bottom panel}: blue-cloud galaxies. 
 Almost all of them are in the star-forming region.
}
\label{fig:iraczall}
\end{figure}

 Figure \ref{fig:iracz} presents the $[3.6]-[8.0]$ colors of these templates 
 as a function of redshift and compares them with those of 
 the \chfour~detections among the red-sequence (red dots) and 
 blue-cloud galaxies (gray scale) with $i<22.5$ at $z<0.7$.
 The IRAS$12112+0305$ template clearly defines the upper edge of the space of 
 the blue-cloud galaxies. Meanwhile, the {\tt Mixed} template 
 follows the lower edge of the blue-cloud galaxies. Most of the red-sequence 
 galaxies have redder colors than the typical SSP template, SSP10G04Z, which 
 should not be surprising because very few galaxies are dust-free and can be 
 accurately represented by SSPs.  There are also many red-sequence galaxies 
 scattered above the {\tt Mixed} template, 
 indicating they are in fact actively forming stars but their optical colors
 are reddened due to dust extinction. We also see that above $z=0.5$, 
 the colors of SF and quiescent galaxies start to blend into each other
 because the $6.2~\mu$m PAH emission band shifts out of the \chfour~channel and
 we can not use the $[3.6]-[8.0]$ color to identify SF galaxies any more.
 Also there are very few galaxies below $z=0.1$ because of the small volume
 in the observation cone.  We therefore only focus our analysis on 
 the redshift range $0.1<z<0.5$.

 In Figure \ref{fig:iraczall}, we present the $[3.6]-[8.0]$ colors of all 
 \chfour~detections in the flux-limited sample split by optical color. 
 In the top panel, we show that a significant fraction of the red-sequence 
 galaxies are scattered into the SF region. In the middle panel, we show that 
 most of the green-valley galaxies are actually actively forming stars
 at roughly the same level as the blue-cloud galaxies. Therefore, that
 the green valley is redder than the blue cloud is likely not an
 indication of a much lower SSFR, but may be due to dust reddening.
 In the bottom panel, we show almost all of the blue-cloud galaxies 
 appear in the SF region.

\begin{figure}
\epsscale{1.2}
\plotone{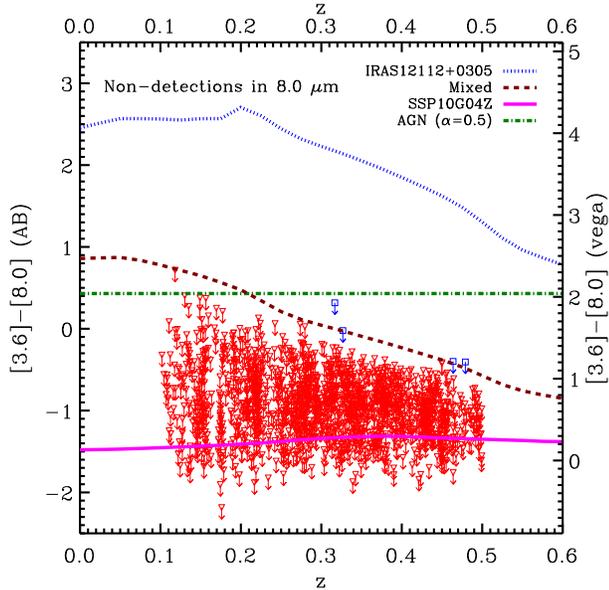}
\caption{IR Color $[3.6]-[8.0]$ vs. z for non-detections in the \chfour~channel 
 in the flux-limited red-sequence sample.
 We use the flux limits $[8.0]<19.9$ ($21.0$) mag in the SWIRE (S-COSMOS)
 fields to calculate the upper limits of their $[3.6]-[8.0]$ colors. 
 Almost all the non-detections fall below the demarcation line 
 (the {\tt Mixed} template).  There are four galaxies that can be 
 identified as star-forming galaxies if their $[8.0]$ fluxes are 
 actually close to the flux limits. We treat them as quiescent galaxies, 
 but identifying them as star-forming galaxies does not change our results.
}
\label{fig:iraczchfour}
\end{figure}

\begin{figure}
\epsscale{1.2}
\plotone{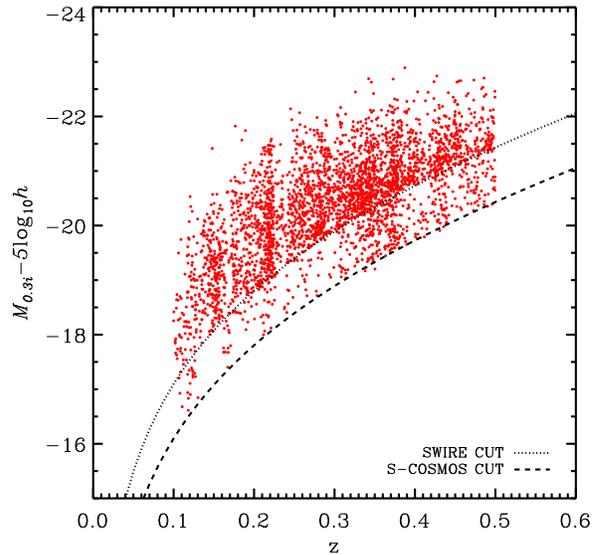}
\caption{\mi~vs. $z$ of the flux-limited red-sequence sample.
 The dotted line shows the flux cut $i=20$ mag in the SWIRE fields and 
 the dashed line represents the cut $i=21$ mag in the S-COSMOS field, 
 corrected with average $K$-corrections of red-sequence galaxies.
 Due to Malmquist bias, we are investigating higher luminosity 
 at higher redshift. 
}
\label{fig:iz}
\end{figure}

\begin{figure}
\epsscale{1.2}
\plotone{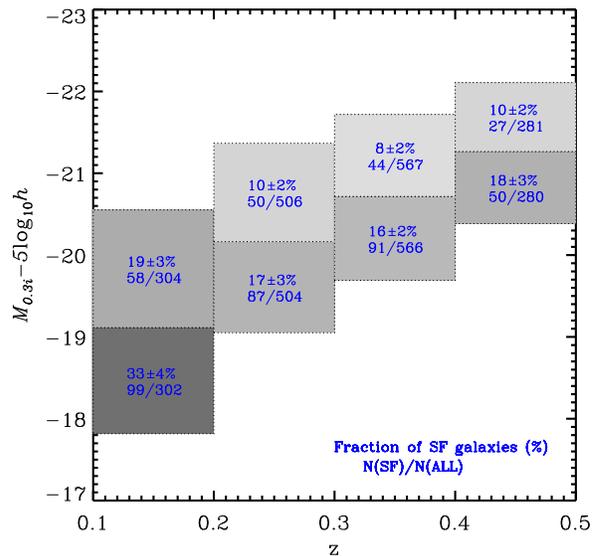}
\caption{Fraction of obscured SF galaxies on the red sequence 
 as a function of luminosity and redshift. In each redshift bin, 
 we divide the sample into two subsamples with equal number of galaxies based 
 on their luminosities and calculate the fraction of SF galaxies and Poisson 
 errors. At all redshifts, the fraction is higher at lower luminosity. We 
 visualize the fraction with gray scale (darker color represents 
 higher fraction). 
}
\label{fig:lumgray}
\end{figure}

\begin{deluxetable*}{ccccccc}
\tabletypesize{\scriptsize}
\tablecolumns{12}
\tablecaption{Fraction of obscured star-forming galaxies on the red sequence}
\tablehead{
\colhead{z} & \colhead{$\langle{M_{^{0.3}i}-5\log_{10}h}\rangle$} &  \colhead{$\langle{M_{^{0.3}i}-5\log_{10}h}\rangle_{\rm cor}$\tablenotemark{a}}
& \colhead{$N_{\rm RS}$\tablenotemark{b}} &  \colhead{$N_{\rm SF}$\tablenotemark{c}} & \colhead{Fraction of SF galaxies}}
\startdata   
$0.1<z<0.2$               & $-18.47$ & $-18.64$ & $302$ & $99$  & $33\pm 4\%$ \\
$\langle{z}\rangle=0.16$  & $-19.83$ & $-20.01$ & $304$ & $58$  & $19\pm 3\%$ \\
\\
$0.2<z<0.3$               & $-19.61$ & $-19.67$ & $504$ & $87$ & $17\pm 3\%$ \\
$\langle{z}\rangle=0.25$  & $-20.77$ & $-20.83$ & $506$ & $50$  & $10\pm 2\%$ \\
\\
$0.3<z<0.4$               & $-20.20$ & $-20.15$ & $566$ & $91$  & $16\pm 2\%$ \\
$\langle{z}\rangle=0.35$  & $-21.22$ & $-21.16$ & $567$ & $44$  & $ 8\pm 2\%$ \\
\\
$0.4<z<0.5$               & $-20.83$ & $-20.65$ & $280$ & $50$  & $18\pm 3\%$ \\
$\langle{z}\rangle=0.44$  & $-21.69$ & $-21.52$ & $281$ & $27$  & $10\pm 2\%$ \\
%
\enddata
\tablenotetext{a}{$\langle{M_{^{0.3}i}-5\log_{10}h}\rangle$ corrected to $z=0.3$, assuming $dM_{0.3_i}/dz=-1.2$ mag per unit redshift.}
\tablenotetext{b}{Number of red-sequence (RS) galaxies.}
\tablenotetext{c}{Number of star-forming (SF) galaxies in the red-sequence subsample.}
\label{tbl:osffrac}
\end{deluxetable*}

 The classification of SF and quiescent galaxies requires detections in 
 both the \chone~and \chfour~channels. Among the $3310$ red-sequence 
 galaxies in the flux-limited sample, $1288$ of them do not have matched 
 \chfour~detections. For these galaxies, we use the flux limits of 
 $[8.0]<19.9$ ($21.0$) mag in the SWIRE (S-COSMOS) fields, to calculate 
 the upper limits of their $[3.6]-[8.0]$ colors.  We show the results in 
 Figure \ref{fig:iraczchfour}.  Only four of these objects may be identified 
 as SF galaxies if their $[8.0]$ fluxes are close to the flux limits. 
 Therefore the flux cuts in $i$ band we adopt in Section \ref{sec:ilimit} 
 have provided a sample with bright \chone~band fluxes 
 (Figure \ref{fig:chonei}) and enable us to reliably classify almost all 
 of the \chfour~non-detections as quiescent galaxies.  Below we treat the 
 ambiguous four objects as quiescent galaxies as well. However, we emphasize 
 that treating them as SF galaxies does not have any noticeable effect in 
 our analysis. We present the results in the next section.

\section{Results: Obscured star formation on the red sequence}\label{sec:results}

 With the classification scheme of AGNs, SF galaxies, and quiescent galaxies 
 defined in Section \ref{sec:selection}, we present a quantitative study 
 of the fraction of obscured SF galaxies on the red sequence in this section.

\subsection{Overall contribution}

 In total, we have $3310$ red-sequence galaxies in the flux-limited sample
 at $0.1<z<0.5$.
 Among these we identify $15\%$ ($506$) as SF galaxies, 
 $84\%$ ($2783$) as quiescent galaxies, and $1\%$ ($21$) as AGNs. 
 Note if we did not exclude $X$-ray detections in about two-thirds
 of the total $6.7~\sdeg$ fields, we would have $31$ AGNs instead.
 In this flux-limited red-sequence sample, we find that overall 
 $15\%$ have IR colors consistent with star formation, indicating their
 optical colors are likely reddened due to dust extinction.

\subsection{Luminosity dependence and redshift evolution}

 It is well-known that elliptical galaxies dominate the bright end of
 the red sequence while disk-dominated systems dominate the faint end
 \citep[\eg][]{marinoni99a, bundy10a, zhu10a}. If the obscuration of the
 SF galaxies on the red sequence is due to high inclination,
 it is likely that the fraction of obscured SF galaxies also depends 
 on luminosity.  Additionally, the global star formation rate (SFR)
 has declined by roughly an order-of-magnitude since $z\sim1$
 \citep[\eg][]{madau96a, hopkins06a, zhu09a, rujopakarn10a}, 
 therefore we may expect that the fraction of obscured SF galaxies on the 
 red sequence has declined with decreasing redshift.  
 We investigate the luminosity dependence and redshift evolution in this section.

\begin{figure}
\epsscale{1.2}
\plotone{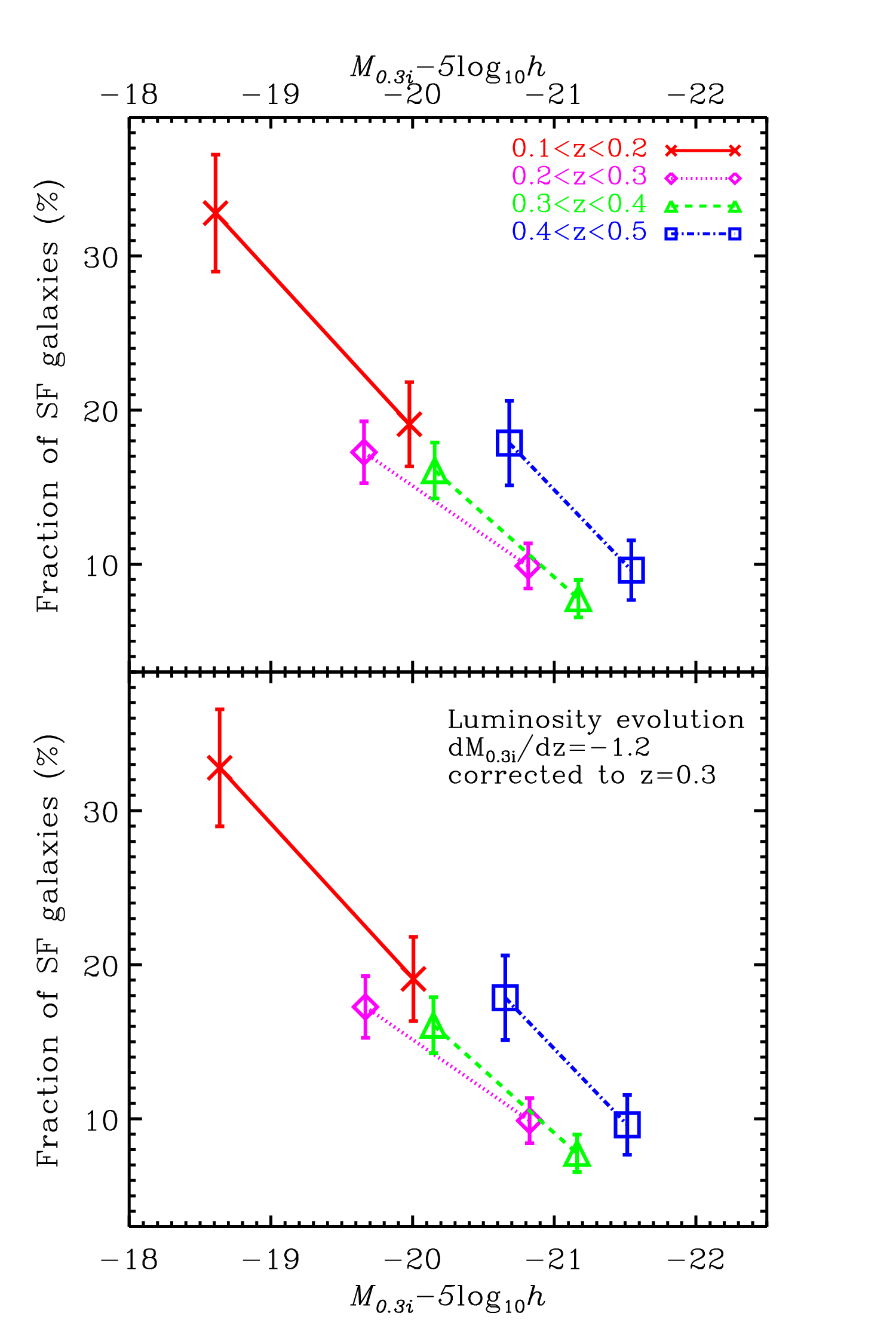}
\caption{Fraction of obscured SF galaxies on the red sequence 
 as a function of luminosity and redshift, as in Figure \ref{fig:lumgray}.  
 We show the fraction at the mean magnitude in each 
 subsample.  In the bottom panel, we assume that $^{0.3}i$ band absolute 
 magnitude fades by $1.2$ per unit redshift and correct all magnitudes 
 to $z=0.3$.  In all redshift bins, the fraction (in percent) is higher 
 at lower luminosity by $\sim 8\%$ per mag.  
 Assuming linear luminosity dependence, at $L \gtrsim L^{\ast}$, the percentage
 of obscured SF galaxies has decreased by $\sim6\pm3\%$ from $z=0.44$ 
 to $z=0.25$.  At $L \lesssim L^{\ast}$, the data from $z=0.35$ to $z=0.16$ 
 are consistent with each other, within the errors. 
 At fainter luminosity ($\gtrsim 1$ mag fainter than $L^{\ast}$), 
 we do not have a wide enough redshift baseline to test for redshift evolution.
}
\label{fig:lumfrac}
\end{figure}

 For a flux-limited sample, due to Malmquist bias, the average luminosity 
 is higher at higher redshift. We show this in Figure \ref{fig:iz}, where we
 plot the absolute magnitude \mi~versus $z$ of the red-sequence galaxies.
 We also show the $i$ band flux cuts, $20$ mag in the SWIRE fields and $21$ mag
 in the S-COSMOS field, corrected with average $K$-corrections of red-sequence 
 galaxies, with the dotted and dashed lines, respectively.  

 To overcome the Malmquist bias and possible degeneracy 
 between luminosity and redshift dependence,
 we therefore define four redshift bins with a small binsize $\Delta z=0.1$. 
 In each redshift bin, we divide the sample into two subsamples with equal 
 number of red-sequence galaxies based on their \mi~magnitudes. 

 Figure \ref{fig:lumgray} presents the fraction of SF galaxies with 
 Poisson errors at the mean absolute magnitude and redshift of each subsample. 
 We visualize the fraction with gray scale (darker colors indicate 
 higher fractions).  We also list the results in Table \ref{tbl:osffrac}.
 The fraction appears to be higher at lower luminosity in all four redshift 
 bins.

 We show the luminosity dependence in Figure \ref{fig:lumfrac}, 
 where we plot the fraction as a function of the mean \mi~in each subsample 
 in the top panel.  In the bottom panel, we assume that the luminosity fades 
 with decreasing redshift by $1.2$ mag per unit redshift, which is the 
 luminosity evolution factor for an SSP with \citet{salpeter55a} IMF formed 
 at $z\sim4$ \citep[\eg][]{vandokkum08a}, 
 and correct all luminosities to $z=0.3$. 
 We find that on average, at $0.1<z<0.5$, the fraction of SF galaxies 
 on the red sequence is about $15\%$ at $\mi\sim-20.5$ 
 (about $L^{\ast}$\footnote{At $z\sim0.1$, the $L^{\ast}$ 
 is $\mr\sim-20.4$ in $^{0.1}r$ band \citep{blanton03c} and 
 $^{0.1}r$ band is very close to $^{0.3}i$ band.}).
 And at all redshifts, the fraction (in percent) increases by 
 $\sim 8\%$ per mag with decreasing luminosity to $0.2L^\ast$. For
 example, in the redshift range $0.3<z<0.4$, the fraction increases
 from 
 $8\pm2\%$ to $16\pm2\%$ 
as luminosity decreases from $\mi\sim-21.2$ 
to $-20.2$.

 Also due to Malmquist bias, at different luminosities we are investigating 
 different redshift ranges. \citet{bundy10a} find that at stellar mass 
 $\lesssim10^{11} M_\odot$, the abundance of red-sequence disk galaxies 
 increases with decreasing redshift, but not as fast as red-sequence 
 spheroidals.
 They also find that at higher stellar mass, the disk population has 
 declined since $z\sim1$. 
 If the obscured SF population behaves as their selected disk population, 
 we expect that the redshift evolution of the fraction also depends on luminosity. 
 With our flux-limited sample, each luminosity range covers a different
 redshift range, complicating our attempt to investigate this question.
 Figure \ref{fig:lumfrac} does not reveal any strong trend of evolution 
 strength with luminosity, but we cannot address this question substantially
 below $L^\ast$ with this sample.
 With our sample, assuming linear luminosity dependence, 
 at $L \gtrsim L^{\ast}$ (\eg at $\mi\sim-21$), the percentage of obscured 
 SF galaxies on the red sequence has decreased by $\sim6\pm3\%$ from $z=0.44$ 
 to $z=0.25$.  At $L \lesssim L^{\ast}$, the data from $z=0.35$ to $z=0.16$ are 
 consistent with each other, within the errors.
 At fainter luminosity ($\gtrsim 1$ mag fainter than $L^{\ast}$), 
 we do not have a wide enough redshift baseline to put interesting
 constraints on evolution.

\subsection{The origin of $24~\mu$m emission on the red sequence}

\begin{figure}
\epsscale{1.2}
\plotone{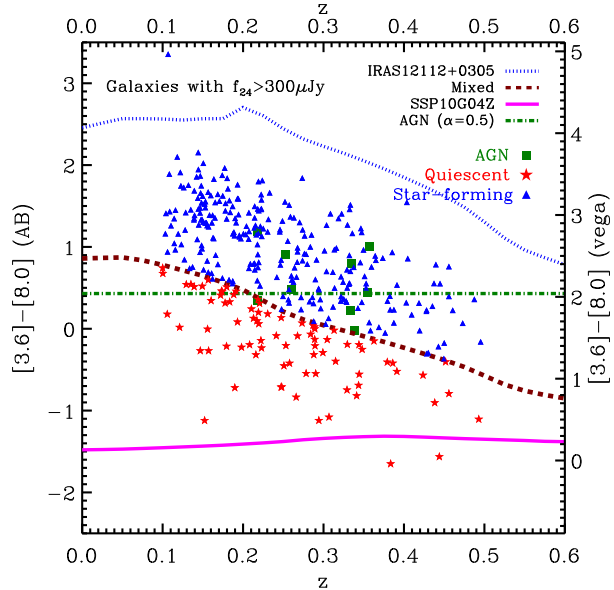}
\caption{IR color $[3.6] - [8.0]$ vs. $z$ for red-sequence galaxies with 
\mipschone~emission $f_{24}>300~\mu$Jy. We show star-forming galaxies in blue 
filled triangles, quiescent galaxies in red filled stars, and AGNs in green 
filled squares.  At $0.1<z<0.5$, there are $355$ ($\sim11\%$ of 
the sample) galaxies with $f_{24}>300~\mu$Jy, among which we identify 
$\sim74\%$ ($265$) as SF galaxies, $\sim23\%$ ($81$) as quiescent galaxies, 
and $\sim3\%$ ($9$) as AGNs. The various lines represent spectral templates, 
as in Figure \ref{fig:template}.
}
\label{fig:excessmips}
\end{figure}
 
 We have shown that a significant fraction ($15\%$) of red-sequence galaxies 
 have IR colors consistent with star formation.  Another approach to search 
 for star formation is to use the \mipschone~band of MIPS, also onboard 
 {\it Spitzer}.  
 The \mipschone~emission is a close tracer of SFR of SF galaxies 
 \citep[\eg][]{calzetti07a, rieke09a}.
 Previous studies \citep[\eg][]{davoodi06a, rodighiero07a, brand09a}
 find $\sim10-20\%$ of red-sequence galaxies with \mipschone~emission 
 $f_{24}\gtrsim300~\mu$Jy.
 With the \mipschone~imaging data, we also find that in our flux-limited 
 red-sequence sample at $0.1<z<0.5$, about $11\%$ ($355$) have \mipschone~flux
 $f_{24}>300~\mu$Jy. At $z=0.3$, this \mipschone~flux corresponds to
 a SFR $\sim 3~M_{\odot}{\rm yr}^{-1}$ \citep{rieke09a}.

 The \mipschone~emission, however, can also originate from AGNs 
 \citep[\eg][]{farrah03a, brand06a}.  
 It is interesting to investigate whether star formation or AGN or 
 both activities are responsible for the \mipschone~emission of these 
 red-sequence galaxies. 
 Previous studies \citep[\eg][]{davoodi06a, rodighiero07a, brand09a} 
 find that both star formation and AGN activities could be
 responsible.
 \citet{brand09a} select $89$ ($\sim10\%$) red-sequence galaxies with 
 $f_{24} > 300~\mu$Jy from the AGN and Galaxy Evolution Survey (AGES) and 
 employ both the emission-line diagnostic diagram \citep[][BPT]{baldwin81a} 
 and the IRAC color-color diagram to investigate the origin of the emission.
 They find that in their sample, with the former method the majority 
 ($\sim 64\%$) are identified as AGNs, while with the latter they identify 
 $\sim 64\%$ as SF galaxies, $\sim31\%$ as quiescent galaxies, 
 and only $\sim5\%$ as AGNs. 
 Therefore both star formation and AGN activities take place in their sample,
 but in the optical the star formation is obscured by dust and in the IR 
 the AGN emission is outshone by star formation emission.

 The PRIMUS spectroscopy does not resolve the narrow lines required to construct
 the BPT diagram. However, we can perform the same AGN,
 SF galaxy, and quiescent galaxy classification as above using IRAC colors.
 Among the $355$ red-sequence galaxies with $f_{24}>300~\mu$Jy, we identify 
 $\sim74\%$ ($265$) as SF galaxies, $\sim23\%$ ($81$) as quiescent 
 galaxies, and $\sim3\%$ ($9$) as AGNs.  
 In Figure \ref{fig:excessmips}, we plot the $[3.6]-[8.0]$ colors of these 
 galaxies as a function of redshift.
 Our results are consistent with \citet{brand09a} and suggest that star 
 formation should account for at least part of the \mipschone~emission.

\section{Discussion}\label{sec:discussion}

\subsection{The nature of galaxies on the red sequence}

 Although the red sequence is a well-defined locus with very small scatter 
 in the color-magnitude diagram, it actually consists of both early-type and 
 late-type galaxies: early-type spheroidal galaxies, late-type disk-dominated
 galaxies without ongoing star formation, and late-type dusty SF galaxies with 
 their optical colors reddened due to dust extinction (\eg edge-on Sb/Sc, etc.).
 When studying galaxy evolution based on optical color, it is therefore 
 necessary to understand the contribution of different types of galaxies on 
 the red sequence. Here we describe a number of methods of doing so,
 finding that broadly speaking they agree with our results here on
 the fraction of star-forming and/or disk galaxies on the red sequence.

 There are many ways to study the composition of galaxies on the red sequence.
 A non-exhaustive list includes: 
 \begin{itemize}\addtolength{\itemsep}{-0.5\baselineskip}
 \item[(1)] direct classification into different morphologies 
 \citep[\eg][]{marinoni99a, yan03a, 
 moustakas04a, bell04a, weiner05a, lotz08a, blanton09a, bundy10a, zhu10a};
 \item[(2)] using inclination-corrected properties 
 \citep[\eg][]{tully98a, masters03a, shao07a, driver07a, bailin08a, 
 bailin08b, unterborn08a, maller09a};
 \item[(3)] SED fitting where dust extinction is a free parameter
 \citep[\eg][]{kauffmann03a, wolf04a, brammer09a};
 \item[(4)] using optical color and optical-near-IR (NIR) color-color diagram
 \citep[\eg][]{pozzetti00a, labbe05a, wuyts07a, williams09a};
 \item[(5)] using IR color  
 \citep[\eg][this work]{brand09a};
 \item[(6)] using MIR flux
 \citep[\eg][this work]{coia05a, davoodi06a, rodighiero07a, saintonge08a, brand09a};
 \item[(7)] using emission line measurements
 \citep[\eg][]{hammer97a, demarco05a, popesso07a, verdugo08a}.
 \end{itemize}

 It is difficult to carry out an exact comparison of these studies because of 
 different sample selection methods, different luminosity and mass ranges 
 investigated, and different redshift ranges investigated. However, all of these
 studies lead to the conclusion that a significant fraction of red-sequence 
 galaxies are disk galaxies and/or obscured SF galaxies. Detailed studies 
 conclude that the fraction depends strongly on luminosity and/or mass.
 In terms of morphology, in the local universe, various papers
 \citep[\eg][]{marinoni99a, zhu10a} have shown that 
 spheroidal galaxies dominate the bright end ($\gtrsim L^{\ast}$) and 
 disk-dominated systems dominate at the faint end.
 At redshifts $0.2<z<1.2$, \citet{bundy10a} find that disk galaxies 
 represent nearly one-half of all red-sequence population and dominate at 
 low stellar mass $\lesssim10^{10}~M_\odot$.
 In terms of star formation, we find at redshifts $0.1<z<0.5$, the fraction of
 obscured SF galaxies on the red sequence increases by $\sim8\%$ per mag
 with decreasing luminosity to $L\sim0.2L^\ast$.

 The redshift evolution of the relative fractions of different components on the 
 red sequence, however, is less clear.
 \citet[][see also \citealt{yan03a}]{moustakas04a} find that the fraction of 
 spheroidal galaxies in extremely red objects (EROs) is $33\%-44\%$ 
 at redshift $\sim1-2$. \citet{lotz08a} find that at luminosity brighter 
 than $0.4L_{B}^{\ast}$, edge-on and dusty disk galaxies are almost one 
 third of the red sequence at $z\sim1.1$, while they only make up 
 $\lesssim10\%$ at $z\sim0.3$.  We also find that at $L \gtrsim L^\ast$, the 
 fraction of obscured SF galaxies decreases by $\sim6\pm3\%$ 
 from $z=0.44$ to $z=0.25$.
 However, other studies do not find a noticeable redshift dependence. 
 \citet{bell04a} and \citet{weiner05a} find that the fraction 
 of edge-on disk galaxies is similar at $z\sim1$ and at $z\sim0$.
 One possible reason for the disagreement is that the redshift evolution of 
 relative fractions may depend on other properties such as stellar mass or 
 luminosity.  For example, \citet{bundy10a} find that at stellar mass 
 $\lesssim10^{11} M_\odot$, the abundance of red-sequence disk galaxies 
 increases with decreasing redshift, but not as fast as red-sequence 
 spheroidals; meanwhile, at higher stellar mass, the disk population has declined 
 since $z\sim1$.

 The relative fractions of different types of galaxies 
 and their redshift evolution
 on the red sequence may also depend on environment.  It is well-known that 
 overall early-type galaxies are more frequently found in high-density 
 environments while late-type galaxies tend to dominate in low-density
 regions \citep[the Morphology-density relation,][]{dressler80a}. 
 There is also the SSFR-density relation such that the mean SSFR is higher 
 at low density, an effect that is important in the local universe as
 well as
 at $z\sim1$ 
 \citep[\eg][]{kauffmann04a, cooper08a}.  Meanwhile, the mean SFR is higher 
 at low density in the local universe but the SFR-density
 relation appears to be reversed at $z\sim1$ \citep[\eg][]{elbaz07a, 
 cooper08a}.  The fraction of obscured SF galaxies therefore may 
 also depend on environment \citep[\eg][]{wolf05a, gallazzi08a, 
 wolf09a, feruglio10a}.
 These studies suggest that in interpreting the 
 optical color-magnitude diagram as a function of environment, the
 population of obscured SF galaxies may need to be accounted for.

%
\begin{figure*}
\epsscale{1.0}
\plotone{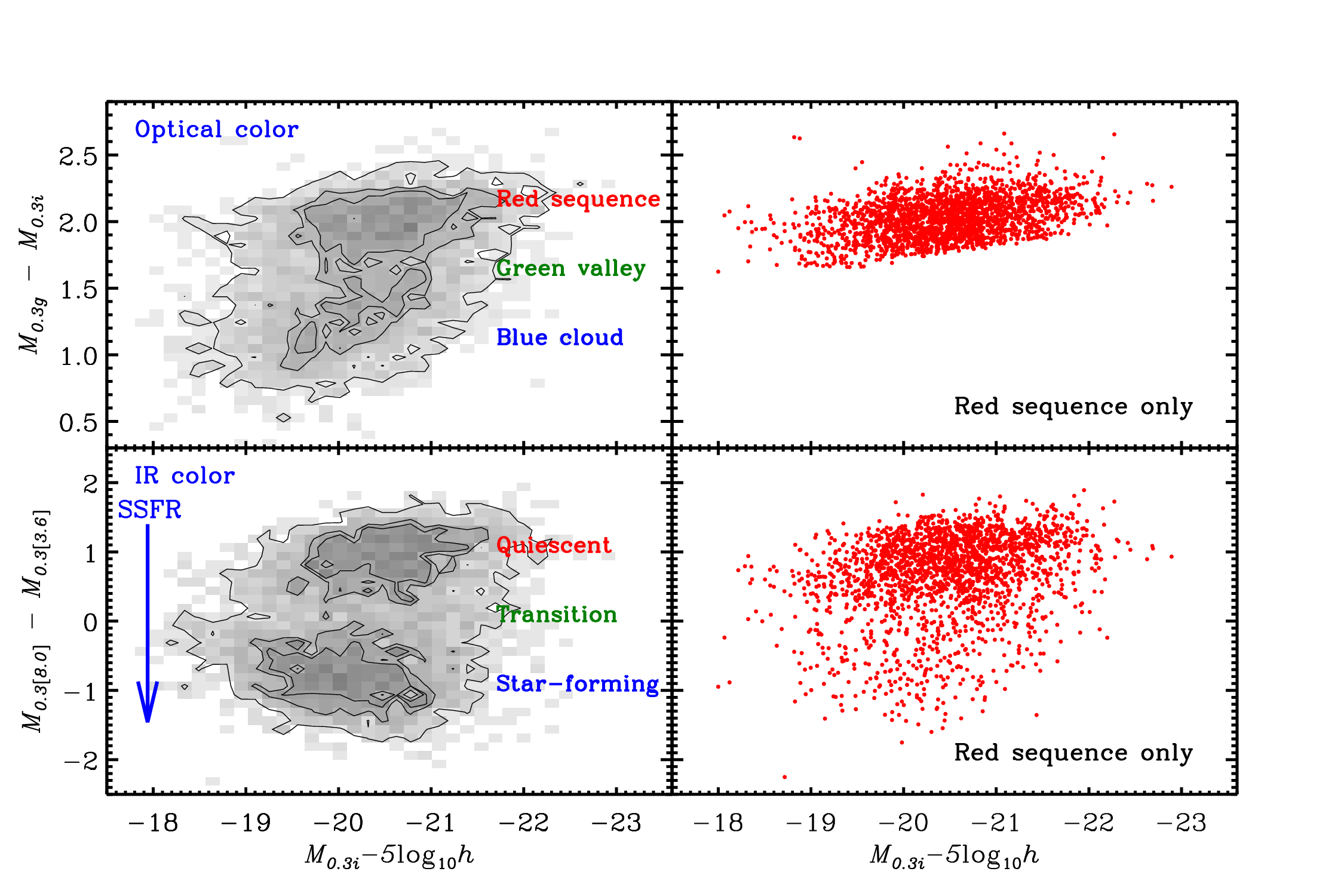}
\caption{Color-magnitude diagram for galaxies between $z=0.2$ and $z=0.4$ in 
 the flux-limited sample.  {\it Upper-left panel}: 
 Optical color \mgmi~vs. $\mi$.
 {\it Upper-right panel}: Optical color \mgmi~vs. $\mi$, but only for 
 red-sequence galaxies.
 {\it Lower-left panel}: IR color \monemtwo~vs. $\mi$.
 {\it Lower-right panel}: IR color  \monemtwo~vs. $\mi$, but only for 
 red-sequence galaxies.  The bimodality is more striking using IR color. 
 And a significant fraction of red-sequence galaxies are scattered into the 
 star-forming region in the IR color-magnitude diagram.
}
\label{fig:newcmd}
\end{figure*}

 An interesting suggestion by some papers is that the obscured SF 
 galaxies on the red sequence are a third type other than 
 red and blue galaxies \citep[\eg][]{wolf05a, weinmann06a, popesso07a, 
 bailin08a, verdugo08a, wolf09a}.
 For example, \citet{wolf05a} and \citet{wolf09a} find these galaxies are 
 distributed differently in environment than both the red and the blue 
 galaxy populations: unlike red galaxies they tend to avoid dense cluster 
 regions, but in contrast to blue galaxies they are not common in 
 low-density fields.  With \mipschone~MIR data, \citet{wolf09a} further 
 show that the mean SSFR of these galaxies in clusters is lower than 
 in the field, in contrast to that of blue galaxies alone, which appears 
 similar in different environments. They also show using {\em HST} imaging
 that edge-on spirals form only a small fraction of spiral galaxies.
 These authors thus suggest that these obscured SF galaxies are a
 transitioning population between blue field galaxies and red cluster 
 galaxies in cores.

 Finally, an interesting question is how much these obscured SF galaxies 
 account for in the overall SFR density in the universe 
 \citep[\eg][]{bell07a}.
 In the flux-limited sample, we have 506 obscured SF galaxies on the red
 sequence, 1451 green-valley and 2497 blue-cloud galaxies.
 Assuming that they have comparable average 
 SSFR and stellar mass, we roughly estimate that of
 order $10\%$ of total cosmic star formation is
 contributed by obscured SF galaxies. An exact analysis
 requires precise SFR measurements, beyond the scope of this paper. 
 We will address this issue in future papers measuring the global SFR 
 density using UV and \mipschone~MIR luminosity.

\subsection{IR color-magnitude diagram}

 As a proxy for the SSFR-stellar mass relation, the optical color-magnitude 
 diagram is a powerful tool in studying galaxy evolution; however, we
 argue here that IR color can be more effective in separating
 star-forming and quiescent galaxies when it is available.  As shown 
 above, optical color can be a misleading variable for SSFR and there is a 
 significant fraction of obscured SF galaxies on the red sequence.
 The fraction depends on luminosity and redshift and may also depend on 
 environment. This complication affects galaxy 
 evolution studies based on optical color. Therefore, it is useful to find more
 suitable variables tracing SSFR.  Examples of such variables are 
 inclination-corrected color \citep[\eg][]{maller09a} and
 extinction-corrected color from SED fitting \citep[\eg][]{brammer09a}.
 IR color is another, and better, variable than optical color to represent
 SSFR; therefore, replacing optical color with IR color when available
 is
 a better strategy when constructing the color-magnitude diagram.

 As an example, in Figure 13 we show both the optical color-magnitude diagram 
 ({\it upper-left panel}) and IR color-magnitude diagram 
 ({\it lower-left panel}), 
 where we have reversed the IR color $[3.6]-[8.0]$ to 
 $[8.0]-[3.6]$ so that SF galaxies are on the bottom.
 We show all the galaxies in the flux-limited sample at $0.2<z<0.4$.
 For non-detections in $8.0~\mu$m channel, we use the lower limits 
 by applying the flux limits $[8.0]<19.9$ ($21.0$) mag 
 in the SWIRE (S-COSMOS) fields.
 We have also applied simple $K$-corrections using the templates 
 for IRAS$12112+0305$ and SSP10G04Z to correct the IR color 
 $[8.0]-[3.6]$ so that it is the color in the channels shifted blueward
 to $z=0.3$, i.e., \monemtwo.
 The bimodality in the IR color-magnitude diagram is more striking than in 
 the optical color-magnitude diagram. This should not be surprising because 
 the bimodality is caused by the short time-scale of transition from star 
 formation to quiescence.  In the right panels, we only plot the red-sequence 
 galaxies and show that a significant fraction of them are in the SF region 
 in the IR color-magnitude diagram, as we quantified in detail above.
 
\section{Summary}\label{sec:summary}

 With the recently completed PRIsm MUlti-object Survey (PRIMUS),
 we perform a quantitative study of the fraction of 
 obscured star-forming galaxies on the red sequence at intermediate redshift. 
 PRIMUS targeted fields with existing
 deep infrared (IR) imaging from the IRAC and MIPS instruments onboard the
 {\it Spitzer Space Telescope}, including $5.7~\sdeg$ fields
 from the SWIRE survey and a $1~\sdeg$ field from the S-COSMOS survey.
 With the precise redshifts from PRIMUS, we select
 in these fields an $i$ band flux-limited sample of $3310$ 
 red-sequence galaxies, with $i<20$ mag in the SWIRE fields 
 and $i<21$ mag in the S-COSMOS field,
 at $0.1<z<0.5$.  With the IR imaging data, we classify
 the red-sequence galaxies into AGNs, star-forming galaxies, and
 quiescent galaxies. Our sample is the largest sample at intermediate redshift
 for such a study and we present for the first time a quantitative 
 analysis of the fraction of obscured star-forming galaxies on the red 
 sequence as a function of luminosity.
 
 We find that on average, at $L \sim L^{\ast}$,
 about $15\%$ of red-sequence galaxies have IR colors ($[3.6]-[8.0]$)
 consistent with star-forming galaxies at $0.1<z<0.5$. 
 The percentage of obscured
 star-forming galaxies increases by $\sim8\%$ per mag with decreasing
 luminosity to $L\sim0.2L^\ast$. At $L \gtrsim L^{\ast}$, the fraction has 
 declined by $\sim6\pm3\%$ from $z=0.44$ to $z=0.25$. At $L \lesssim L^{\ast}$,
 the fraction is consistent with no redshift evolution between $z=0.35$ 
 and $z=0.16$.
 
 Using the \mipschone~imaging from MIPS on board {\it Spitzer}, we find
 that $\sim 11\%$ of red-sequence galaxies exhibit \mipschone~emission 
 with $f_{24}>300~\mu$Jy at redshifts $0.1<z<0.5$, in agreement with 
 previous studies.  With the IRAC imaging, we find that $\sim74\%$ of 
 these galaxies have IR colors consistent 
 with star-forming galaxies, suggesting that star formation accounts for 
 at least part of the \mipschone~emission.

 Our analysis suggests that a significant fraction of red-sequence
 galaxies are actually actively forming stars, but their optical colors
 are reddened due to dust extinction. An accurate understanding of
 galaxy evolution using optical color therefore needs to account for
 this complication. We suggest that IR color, \eg the IRAC color $[3.6]-[8.0]$,
 should be a better color to use when constructing the color-magnitude
 diagram to represent the specific star-formation rate-stellar mass relation.

\acknowledgments

 We acknowledge 
 Rebecca Bernstein, Adam Bolton, Douglas Finkbeiner, David W. Hogg, 
 Timothy McKay, Sam Roweis, and Wiphu Rujopakarn
 for their contributions to the PRIMUS project.
 This paper includes data gathered with the $6.5$ meter Magellan 
 Telescopes located at Las Campanas Observatory, Chile.
 We thank the support staff at LCO for their help during
 our observations, and we acknowledge the use of community access
 through NOAO observing time.  
 Funding for PRIMUS has been provided by NSF grants AST-0607701, 0908246, 
 0908442, 0908354 and NASA grant 08-ADP08-0019.
 RJC is supported by NASA through Hubble Fellowship grant 
 HF-01217 awarded by the Space Telescope Science Institute, 
 which is operated by the Associated of Universities for Research in 
 Astronomy, Inc., for NASA, under contract NAS 5-26555.

 This work is based in part on observations made with the 
 {\it Spitzer Space Telescope}, 
 which is operated by the Jet Propulsion Laboratory (JPL), 
 California Institute of Technology under a contract with NASA.
 We would like to thank the SWIRE and the S-COSMOS teams for 
 making their data publicly available.

\bibliographystyle{apj}
\bibliography{sfrs}


\end{document}